\documentstyle[12pt,axodraw]{article}

\def\beq{\begin{equation}}
\def\eeq{\end{equation}}
\def\beqa{\begin{eqnarray}}
\def\eeqa{\end{eqnarray}}

\newlength{\dinwidth} \newlength{\dinmargin}
\begin{document}
\vspace{2cm}
\begin{center}
{\Large \bf Two-loop and $n$-loop vertex corrections for eikonal diagrams with 
massive partons}
\end{center}
\vspace{2mm}
\begin{center}
{\large Nikolaos Kidonakis}\\
\vspace{2mm}
{\it Department of Physics and Astronomy\\
University of Rochester\\
Rochester, NY 14627-0171} 
\end{center}

\begin{abstract}
We present the ultraviolet poles and finite terms
of two-loop vertex corrections for diagrams with massive partons
in the eikonal approximation. We discover and prove that the results
for a set of the corrections generalize to $n$-loop eikonal diagrams.
These results will enhance theoretical understanding and allow greater 
calculational accuracy for the many partonic processes where the eikonal 
approximation is applied.
\end{abstract}

\thispagestyle{empty} \newpage \setcounter{page}{2}

\pagebreak

Perturbation theory for partonic processes entails the
calculation of increasingly complicated loop diagrams 
as one moves to higher orders. At present, many QCD processes 
and some beyond the Standard Model processes have been
calculated to next-to-leading order (NLO) in the strong coupling
$\alpha_S$, and they require the
evaluation of one-loop diagrams. Higher-order calculations
involve diagrams at two-loops and higher \cite{WGQCD,QCDSM}.

Beyond the use of standard fixed-order calculations in perturbation 
theory, it is possible to evaluate higher-order contributions
to partonic cross sections, that are dominant in certain regions of phase
space, using a panoply of resummation methods and related techniques. 
The eikonal approximation has been an extremely useful
tool in these other approaches to perturbative calculations.
The eikonal approximation is valid for emission 
of soft gluons from partons in the hard scattering
and leads to a simplified form of the Feynman rules. 
When the gluon momentum goes to zero,
the usual Feynman rules for the quark propagator and quark-gluon
vertex in the diagram in Figure 1 simplify as follows:

\beq
{\bar u}(p) \, (-i g_s T_F^c) \, \gamma^{\mu} 
\frac{i (p\!\!/+k\!\!/+m)}{(p+k)^2
-m^2+i\epsilon} \rightarrow {\bar u}(p)\,  g_s T_F^c \, \gamma^{\mu} 
\frac{p\!\!/+m}{2p\cdot k+i\epsilon}
={\bar u}(p)\, g_s T_F^c \,
\frac{v^{\mu}}{v\cdot k+i\epsilon}
\eeq
with $v$ a dimensionless vector, $p \propto v$,  $g_s^2=4 \pi \alpha_s$,
and $T_F^c$ the generators of SU(3) in the fundamental representation.

\vspace*{-1cm}
\begin{figure}[htb]
\begin{center}
\begin{picture}(120,120)(0,0)
\Line(0,80)(0,90)
\Line(3,80)(3,90)
\ArrowLine(0,85)(50,85)
\ArrowLine(50,85)(100,85)
\Vertex(50,85){2}  
\Gluon(50,25)(50,85){2}{8}
\Text(20,75)[c]{$p+k$}
\Text(80,75)[c]{$p$}
\LongArrow(65,50)(65,35)
\Text(85,40)[c]{$ k \rightarrow 0$}
\end{picture}
\end{center}
\vspace*{-1cm}
\caption{\label{eikonal}  Eikonal approximation}
\end{figure}
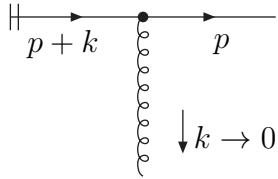

The eikonal approximation is an essential element in calculations
of QCD cross sections in certain kinematical regions.
Apart from its use in deriving the soft part
of the cross section in standard fixed-order QCD calculations, 
it has been applied to calculations of the high-energy behavior of elastic
quark-quark scattering, (near) forward scattering amplitudes, 
and wide-angle elastic scattering \cite{BotSt,SoSt,Korchem,KaKtStW}. 
In addition, the eikonal approximation is invaluable in the development of
transverse momentum and threshold resummations for a variety of 
partonic processes [1, 2, 7-14].

The ultraviolet (UV) pole structure of $n$-loop vertex correction
diagrams involving eikonal lines is particularly worthy of attention, 
because, in addition to its 
intrinsic theoretical interest, it plays a direct role in renormalization 
group evolution equations that have been used in studies of hard 
elastic scattering  \cite{BotSt,SoSt} and 
threshold resummations [10-14]. 
At present, results are known to one loop and have 
been used in state-of-the-art next-to-leading logarithm (NLL) resummations 
and their expansions to NNLO and higher orders for many QCD processes [10-17]. 
Threshold resummations will also play a major role in calculations
of cross sections for Higgs production and supersymmetric particles 
since these particles are expected to be discovered close to threshold.
The successes of NLL resummations, including dramatic reductions in
scale dependence, have been quite remarkable; 
however there are still sizable uncertainties in the calculations
due to formally subleading terms which in practice can have
significant contributions.
To go beyond the level of NLL accuracy and control subleading terms,
two loop and even higher-order eikonal calculations are needed.
These calculations are also useful for increasing theoretical
accuracy in the other many applications of the eikonal approximation,
especially when renormalization group evolution equations are used. 
The calculation of two-loop vertex corrections with eikonal lines
representing massive partons will be the subject of this letter.
The result for the two-loop vertex correction will suggest a generalization
for the form of a set of the corrections at arbitrary $n$ loops.
We conjecture and then prove this $n$-loop result by induction.
Results for massless partons will be presented in future work.

We begin by presenting one-loop vertex corrections in the eikonal approximation
using the axial gauge. A representative one-loop vertex correction
diagram is given in Figure 2.
The general axial gauge gluon propagator is
\begin{equation}
D^{\mu \nu}(k)=\frac{-i}{k^2+i\epsilon} N^{\mu \nu}(k), \quad
N^{\mu \nu}(k)=g^{\mu \nu}-\frac{n^{\mu}k^{\nu}+k^{\mu}n^{\nu}}{n \cdot k}
+n^2\frac{k^{\mu}k^{\nu}}{(n \cdot k)^2},
\end{equation}
with $n^{\mu}$ the axial gauge-fixing vector.
The propagator for a quark, antiquark, or gluon eikonal line is
$i/(\delta v \cdot k +i \epsilon)$ with $\delta=+1 (-1)$ when the 
momentum $k$ flows in the same (opposite) direction as the dimensionless
vector $v$. The interaction gluon-eikonal line vertex for a quark or 
antiquark eikonal line
is $-i \, g_s \, T_F^c \, v^{\mu} \, \Delta$ with $\Delta=+1(-1)$ for a quark
(antiquark). 

\begin{figure}[htb]
\begin{center}
\begin{picture}(120,120)(0,0)
\Vertex(0,50){5}
\ArrowLine(0,50)(60,80)
\ArrowLine(60,80)(100,100)
\Vertex(60,80){2}  
\Gluon(60,80)(60,20){2}{8}
\Text(25,78)[c]{$p_i+k$}
\Text(88,100)[c]{$p_i$}
\LongArrow(70,55)(70,45)
\Text(80,50)[c]{$k$}
\ArrowLine(0,50)(60,20)
\ArrowLine(60,20)(100,0)
\Vertex(60,20){2}
\Text(25,20)[c]{$p_j-k$}
\Text(88,0)[c]{$p_j$}
\end{picture}
\end{center}
\caption{\label{oneloop}  One-loop vertex correction diagram}
\end{figure}
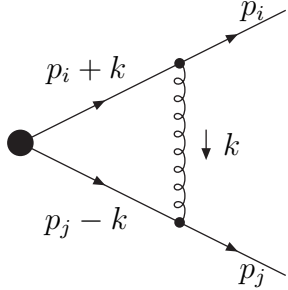

We denote the kinematic (i.e. color-independent) part of the $n$-loop 
correction to the vertex, with the virtual gluon linking lines 
$i$ and $j$, as $\omega^{(n)}_{ij}(v_{i},v_{j})$.
The eikonal lines represent massive partons with mass $m$, for example
a heavy quark-antiquark pair.
The one-loop expression for $\omega_{ij}$ is then
\beq
\omega_{ij}^{(1)}(v_{i},v_{j})\equiv
{g}_{s}^2\int\frac{d^D k}{(2\pi)^D}\frac{-i}{k^2+i\epsilon}
N^{\mu \nu}(k) \frac{\Delta_{i} \: v_{i}^{\mu}}
{\delta_{i} v_{i} \cdot k+i\epsilon}\;
\frac{\Delta_{j} \:v_{j}^{\nu}} {\delta_{j}v_{j} \cdot k+i\epsilon}\, .
\eeq
After isolating the UV poles of the integral in dimensional
regularization (with $\varepsilon=4-D$) using a variety of techniques,  
we can present the UV-pole part of $\omega_{ij}^{(1)}$ as \cite{KS}
\beq       
\omega_{ij}^{(1)\; \rm{UV}}(v_{i},v_{j})=
{\cal S}_{ij}^{(1)} \, \frac{\alpha_{s}}{\pi\varepsilon}
\left[L_{\beta} + L_i + L_j -1 \right] \ ,
\label{omegaheavy}
\eeq
where ${\cal S}_{ij}^{(1)}$ is an  overall sign,
${\cal S}_{ij}^{(1)}=\Delta_i \: \Delta_j \: \delta_i \: \delta_j$,
$L_\beta$ is the velocity-dependent eikonal function
\begin{equation}
L_{\beta}=\frac{1-2m^2/s}{\beta}\left[\ln\left(\frac{1-\beta}{1+\beta}
\right)+\pi i \right]\, ,
\label{Lb}
\end{equation}
with $\beta=\sqrt{1-4m^2/s}$, $s=(p_i+p_j)^2$, and 
the $L_i$ and $L_j$ are rather complicated functions of
the gauge vector $n$. We note that these gauge-dependent functions
are cancelled by the inclusion of contributions from 
one-loop self-energy corrections \cite{KS}.

We now continue with a calculation of the two-loop diagram
in Figure 3.

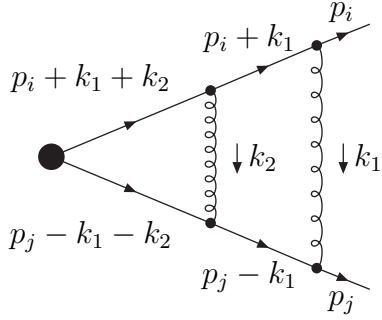
\begin{figure}[htb]
\begin{center}
\begin{picture}(120,120)(0,0)
\Vertex(0,60){5}
\ArrowLine(0,60)(60,85)
\ArrowLine(60,85)(100,102)
\ArrowLine(100,102)(120,110)
\Vertex(60,85){2}
\Vertex(60,35){2}  
\Gluon(60,85)(60,35){2}{8}
\Text(15,90)[c]{$p_i+k_1+k_2$}
\Text(75,105)[c]{$p_i+k_1$}
\Text(110,115)[c]{$p_i$}
\LongArrow(70,65)(70,55)
\Text(80,60)[c]{$k_2$}
\ArrowLine(0,60)(60,35)
\ArrowLine(60,35)(100,18)
\ArrowLine(100,18)(120,10)
\Vertex(100,102){2}
\Vertex(100,18){2}
\Gluon(100,102)(100,18){2}{8}
\Text(15,30)[c]{$p_j-k_1-k_2$}
\Text(75,15)[c]{$p_j-k_1$}
\Text(110,5)[c]{$p_j$}
\LongArrow(110,65)(110,55)
\Text(120,60)[c]{$k_1$}
\end{picture}
\end{center}
\caption{\label{twoloop}  Two-loop vertex correction diagram}
\end{figure}

At two loops, the expression for $\omega_{ij}$ is 
\beqa   
\omega_{ij}^{(2)}(v_{i},v_{j})&=&
{g}_{s}^4 \int\frac{d^D k_1}{(2\pi)^D} \, \frac{(-i)}{k_1^2+i\epsilon} \,
N^{\mu \nu}(k_1)
\int\frac{d^D k_2}{(2\pi)^D} \, \frac{(-i)}{k_2^2+i\epsilon} \,
N^{\rho \sigma}(k_2)
\nonumber \\ && \hspace{-30mm} \times \,
\frac{\Delta_{1i} \: v_{i}^{\mu}}
{\delta_{1i} v_{i} \cdot k_1+i\epsilon} \;
\frac{\Delta_{2i} \: v_{i}^{\rho}}
{\delta_{2i} v_{i} \cdot (k_1+k_2)+i\epsilon} \;
\frac{\Delta_{2j} \:v_{j}^{\sigma}} 
{\delta_{2j}v_{j} \cdot (k_1+k_2)+i\epsilon} \; 
\frac{\Delta_{1j} \:v_{j}^{\nu}} 
{\delta_{1j}v_{j} \cdot k_1+i\epsilon} \, .
\label{omega2}
\eeqa
As for the one-loop calculation, 
the result for $\omega_{ij}^{(2)}$ depends on whether 
the eikonal lines represent massless or massive partons. 
For massive eikonal lines, that we discuss here,
the highest UV poles encountered are $1/\varepsilon^2$.

We may rewrite Eq. (\ref{omega2}) as
\beq
\omega^{(2)}_{ij}(v_{i},v_{j})= 
{\cal S}_{ij}^{(2)}
\sum_{k,l=1,2,3} I_{kl}^{(2)}(v_i, v_j) \, ,
\label{omegaII}
\eeq
where $I_{kl}^{(2)}$ denotes the contribution of the 
$k$-th term in the gluon propagator $N^{\mu \nu}(k_1)$ in Eq. (2)
with the $l$-th term in $N^{\rho \sigma}(k_2)$,
and ${\cal S}_{ij}^{(2)}$ is an  overall sign,
${\cal S}_{ij}^{(2)}=\Delta_{1i} \: \Delta_{1j} \: \Delta_{2i} \: 
\Delta_{2j} \: \delta_{1i} \: \delta_{1j} \:
\delta_{2i} \: \delta_{2j}$.

For example, the integral $I_{11}^{(2)}$ is defined as
\beqa
I_{11}^{(2)}(v_i,v_j) &\equiv& g_s^4 \int \frac{d^D k_1}{(2\pi)^D} \, 
\frac{(-i)}{k_1^2+i\epsilon} \; \frac{v_i \cdot v_j}
{(v_i \cdot k_1+i\epsilon) (v_j \cdot k_1+i\epsilon)}
\nonumber \\ && \quad\quad \times 
\int \frac{d^D k_2}{(2\pi)^D} \, 
\frac{(-i)}{k_2^2+i\epsilon} \; \frac{v_i \cdot v_j}
{[v_i \cdot (k_1+k_2)+i\epsilon] [v_j \cdot (k_1+k_2)+i\epsilon]}
\nonumber \\ && \hspace{-23mm}
=g_s^4 \int \frac{d^D k_1}{(2\pi)^D} \, 
\frac{i\, \pi^{\varepsilon/2} 2^{2\varepsilon} \, \Gamma(1+\varepsilon/2)\,
(1-2m^2/s)^2}{4 \pi^2 \; (k_1^2+i\epsilon) \; 
(v_i \cdot k_1+i\epsilon) \; (v_j \cdot k_1+i\epsilon)}
\left[\frac{1}{\varepsilon} F(x)
+\int_0^1 dx \int_0^1 dz \frac{f(x,z)}{(1-z)_+} \right] ,
\label{I11}
\eeqa
where we used $v_i \cdot v_j=1-2m^2/s$, with $m$
the mass of the heavy quark. The function $f$ is of the form
\beq
f(x,z)=\left[A(x,z)+B_i(x,z)\,  k_1\cdot v_i
+B_j(x,z) \, k_1\cdot v_j\right]^{-1}
\eeq
and the $k_1$ integral over it is UV finite,
while 
\beq
F(x)=\int_0^1 dx \, f(x,1)=- \frac{L_{\beta}}{1-2m^2/s} 
\eeq
with $L_{\beta}$ defined in Eq. (\ref{Lb}).

After calculating all the integrals in $\omega_{ij}^{(2)}$, we can
present the two-loop UV-pole terms: 
\beqa
&&\omega_{ij}^{(2)\; \rm{UV}}(v_{i},v_{j})=
{\cal S}_{ij}^{(2)} \, \frac{\alpha_{s}^2}{\pi^2}
\frac{1}{\varepsilon^2}
\left[L_{\beta}+L_i + L_j -1 \right]^2
\nonumber \\ &&
{}+{\cal S}_{ij}^{(2)} \, \frac{\alpha_{s}^2}{\pi^2}
\frac{1}{\varepsilon}\left[L_{\beta}+L_i + L_j -1 \right]
\left\{\left[L_{\beta}+L_i + L_j -1\right]
\left[\ln 2+\ln(4\pi)-\gamma_E\right]\right.
\nonumber \\ && \quad \quad \left.
{}+\ln\left(\frac{n^2}{2}\right)+ v_i \cdot n \left(L_i'-\frac{L_i''}{2}
\right)+ v_j \cdot n \left(L_j'-\frac{L_j''}{2}\right)-M_{\beta}
\right\} \, ,
\label{omega2heavy}
\eeqa
where $\gamma_E$ is the Euler constant, $L'$ and $L''$ are 
complicated functions of the gauge vector $n$, and
\beqa
M_{\beta}&=&\frac{(1-2m^2/s)}{\beta}\left[\ln 2 \;
\ln\left(\frac{1+\beta}{1-\beta}\right)
+\frac{1}{2}\ln^2(1-\beta)-\frac{1}{2}\ln^2(1+\beta)
+{\rm Li}_2\left(\frac{1+\beta}{1-\beta}\right)\right.
\nonumber \\ && \quad \quad \left.
{}-{\rm Li}_2\left(\frac{1-\beta}{1+\beta}\right)
-{\rm Li}_2\left(\frac{1+\beta}{2}\right)
+{\rm Li}_2\left(\frac{1-\beta}{2}\right)\right] \, .
\eeqa
We note that at two loops there are several more diagrams, in addition
to Figure 3, that include two-loop self-energy corrections; 
one-loop self-energy
corrections with one-loop vertex corrections as in Figure 2; 
diagrams with three-gluon vertices; and crossed diagrams.
Some of these diagrams are zero, some are UV finite, and some contribute
gauge-dependent terms that cancel against the gauge terms with 
$\ln (n^2)$, $L$, $L'$,
and $L''$ in $\omega_{ij}^{(2)}$. Details will be given elsewhere.

For the discussion to follow, we define 
$\omega_{ij}'^{(n)}\equiv{\omega}_{ij}^{(n)}/{\cal S}_{ij}^{(n)}$
modulo UV-finite integrals encountered as in Eq. (\ref{I11}) and
infrared terms \cite{NKDPF}.
We now note that the coefficient of the leading UV pole, i.e.
the $1/\varepsilon^2$ pole, in $\omega_{ij}'^{(2)}$ is simply the square
of the coefficient of the $1/\varepsilon$ pole in $\omega_{ij}'^{(1)}$.
This fact suggests the possibility that the $1/\varepsilon$ pole
in $\omega_{ij}'^{(2)}$ might be derived by squaring $\omega_{ij}'^{(1)}$
after we calculate the finite pieces in $\omega_{ij}'^{(1)}$.

A calculation of the finite pieces of $\omega_{ij}'^{(1)}$ gives
\beqa
\omega_{ij}'^{(1)\, {\rm finite}}(v_i,v_j)&=&\frac{\alpha_s}{2\pi}
\left\{\left[L_{\beta}+L_i + L_j -1 \right]
\left[\ln 2+\ln(4\pi)-\gamma_E\right]
+\ln\left(\frac{n^2}{2}\right) \right.
\nonumber \\ && \left.
{}+ v_i \cdot n \left(L_i'-\frac{L_i''}{2}
\right)+ v_j \cdot n \left(L_j'-\frac{L_j''}{2}\right)-M_{\beta}\right\} 
\, .
\eeqa
Then
\beq
\omega_{ij}'^{(1)}=\frac{\alpha_s}{\pi\varepsilon}
\left[L_{\beta}+L_i + L_j -1 \right]
+\omega_{ij}'^{(1)\, {\rm finite}} +{\cal O}(\varepsilon) \,.
\eeq
We see then that indeed $\omega_{ij}'^{(2)}
=[\omega_{ij}'^{(1)}]^2$. This holds not only for all the 
UV poles but also for the finite terms in $\omega_{ij}'^{(2)}$, 
as an explicit calculation,  keeping terms of 
${\cal O}(\varepsilon)$ in $\omega_{ij}'^{(1)}$, verifies.

\begin{figure}[htb]
\begin{center}
\begin{picture}(200,200)(0,0)
\Vertex(0,60){5}
\ArrowLine(0,60)(60,80)
\ArrowLine(60,80)(135,105)
\ArrowLine(135,105)(180,120)
\ArrowLine(180,120)(195,125)
\Vertex(60,80){2}
\Vertex(60,40){2}  
\Gluon(60,80)(60,40){2}{8}
\Text(10,90)[c]{$p_i+k_1+\cdots+k_n$}
\Text(145,130)[c]{$p_i+k_1$}
\Text(190,135)[c]{$p_i$}
\LongArrow(70,65)(70,55)
\Text(80,60)[c]{$k_n$}
\ArrowLine(0,60)(60,40)
\ArrowLine(60,40)(135,15)
\ArrowLine(135,15)(180,0)
\ArrowLine(180,0)(195,-5)
\Vertex(135,105){2}
\Vertex(135,15){2}
\Gluon(135,105)(135,15){2}{8}
\Vertex(180,120){2}
\Vertex(180,0){2}
\Gluon(180,120)(180,0){2}{8}
\Text(10,30)[c]{$p_j-k_1-\cdots-k_n$}
\Text(145,-10)[c]{$p_j-k_1$}
\Text(190,-15)[c]{$p_j$}
\Text(110,60)[c]{$\cdots$}
\LongArrow(145,65)(145,55)
\Text(155,60)[c]{$k_2$}
\LongArrow(190,65)(190,55)
\Text(200,60)[c]{$k_1$}
\end{picture}
\end{center}
\caption{\label{nloop}  $n$-loop vertex correction diagram}
\end{figure}
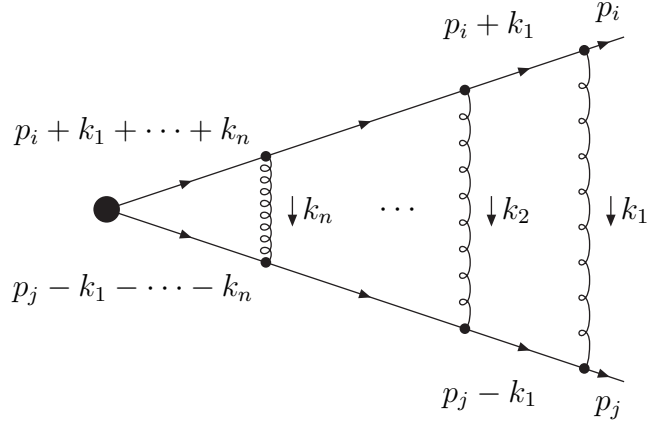

We now propose a conjecture for the generalization of this finding to 
$n$-loops.
The conjecture is that the UV poles and finite terms in the $n$-loop 
vertex correction in Figure 4 can be derived by raising the one-loop
result to the $n$-th power. In other words, we want to prove that
\beq
\omega_{ij}'^{(n)}(v_{i},v_{j})=
\left[\omega_{ij}'^{(1)}(v_{i},v_{j})\right]^n \, .
\eeq
We prove this by induction.
As we have just seen, the above equation holds for $n=2$.
Next, we assume that it holds for arbitrary $n$, and show that
it must then also hold for $n+1$. Now,
\beqa
\omega_{ij}'^{(n+1)}&=& g_s^{2(n+1)} \int \frac{d^D k_1}{(2\pi)^D} \cdots 
\frac{d^D k_n}{(2\pi)^D} \frac{d^D k_{n+1}}{(2\pi)^D}
\nonumber \\ && \times 
\frac{(-i)N^{\mu_1\nu_1}(k_1)}{k_1^2+i\epsilon} \cdots
\frac{(-i)N^{\mu_n\nu_n}(k_n)}{k_n^2+i\epsilon}
\frac{(-i)N^{\mu_{n+1}\nu_{n+1}}(k_{n+1})}{k_{n+1}^2+i\epsilon}
\nonumber \\ && \times 
\frac{v_i^{\mu_1}}{v_i\cdot k_1+i\epsilon} \cdots 
\frac{v_i^{\mu_n}}{v_i\cdot (k_1+\cdots+k_n)+i\epsilon}\;
\frac{v_i^{\mu_{n+1}}}{v_i\cdot (k_1+\cdots+k_n+k_{n+1})+i\epsilon}
\nonumber \\ && \times 
\frac{v_j^{\nu_1}}{v_j\cdot k_1+i\epsilon} \cdots 
\frac{v_j^{\nu_n}}{v_j\cdot (k_1+\cdots+k_n)+i\epsilon}\;
\frac{v_j^{\nu_{n+1}}}{v_j\cdot (k_1+\cdots+k_n+k_{n+1})+i\epsilon} \, .
\eeqa
The $k_{n+1}$ integral gives the same UV poles
and finite terms as $\omega_{ij}'^{(1)}$ plus an integral 
which is UV finite after integration over $k_1 \cdots k_n$.
Therefore, $\omega_{ij}'^{(n+1)}=\omega_{ij}'^{(n)}\omega_{ij}'^{(1)}
=\left[\omega_{ij}'^{(1)}\right]^{n+1}$. 
So our formula also holds for $n+1$. This concludes the proof. 

Thus, we see that the vertex corrections of the form in Figure 4
can be readily derived for any number of loops $n$, by calculating
the one-loop result and keeping terms in it of 
${\cal O}({\varepsilon}^{n-1})$. 
Of course, there is an increasing number of self-energy and other 
graphs as we move to higher $n$, and these graphs are expected to 
provide contributions that will cancel the gauge dependence 
in $\omega_{ij}^{(n)}$.

The theoretical loop calculations outlined in this letter
should provide a powerful tool in the numerous applications
of the eikonal approximation to QCD and beyond, and specifically in 
more accurate resummations for a large variety of QCD and beyond the 
Standard Model partonic processes. Each process has a specific 
partonic content and color structure so the details have to be worked 
separately for each application. But the universal ingredient in all such 
studies is the $n$-loop result presented in this letter.


\begin{thebibliography}{99}

\bibitem{WGQCD}
Summary: Working Group on QCD and Strong Interactions, hep-ph/0201146
and references therein.

\bibitem{QCDSM}
The QCD/SM Working Group: Summary Report, hep-ph/0204316 and
references therein. 

\bibitem{BotSt} 
J. Botts and G. Sterman, Nucl. Phys. {\bf B325}, 62 (1989). 

\bibitem{SoSt}  
M.G. Sotiropoulos and G. Sterman, Nucl. Phys. {\bf B419}, 59 (1994)
and references therein. 

\bibitem{Korchem} 
I.A. Korchemskaya and G.P. Korchemsky, Nucl. Phys. {\bf B437}, 127 (1995) 
and references therein.

\bibitem{KaKtStW}
A.I. Karanikas, C.N. Ktorides, N.G. Stefanis, and S.M.H. Wong, 
Phys. Lett. {\bf B455}, 291 (1999). 

\bibitem{SS}
J.C. Collins and D.E. Soper, Nucl. Phys. {\bf B193}, 381 (1981).

\bibitem{CSS}
J.C. Collins, D.E. Soper, and G. Sterman, Nucl. Phys. {\bf B250}, 199 (1985).

\bibitem{NKOY}
P.M. Nadolsky, N. Kidonakis, F. Olness, and C.-P. Yuan, in DPF2002 Meeting,
hep-ph/0207332.

\bibitem{KS}
N. Kidonakis and G. Sterman, Phys. Lett. B {\bf 387}, 867 (1996);
Nucl. Phys. {\bf B505}, 321 (1997).

\bibitem{NK}
N. Kidonakis, Int. J. Mod. Phys. A {\bf 15}, 1245 (2000).

\bibitem{KOS}
N. Kidonakis, G. Oderda, and G. Sterman,
Nucl. Phys. {\bf B525}, 299 (1998);
{\bf B531}, 365 (1998).

\bibitem{LOS}
E. Laenen, G. Oderda, and G. Sterman,
Phys. Lett. {\bf B438}, 173 (1998).

\bibitem{LSV}
E. Laenen, G. Sterman, and W. Vogelsang,
Phys. Rev. D {\bf 63}, 114018 (2001). 

\bibitem{NK01}
N. Kidonakis, Phys. Rev. D {\bf 64}, 014009 (2001).

\bibitem{KLMV}
N. Kidonakis, E. Laenen, S. Moch, and R. Vogt, 
Phys. Rev. D {\bf 64}, 114001 (2001).

\bibitem{NKJOVD}
N. Kidonakis and J.F. Owens, Phys. Rev. D {\bf 61}, 094004 (2000); 
Phys. Rev. D {\bf 63}, 054019 (2001); 
N. Kidonakis and V. Del Duca, Phys. Lett. {\bf B480}, 87 (2000). 

\bibitem{NKDPF}
N. Kidonakis, in DPF2002 Meeting, hep-ph/0207142.

\end{thebibliography}
\end{document}